\documentclass[useAMS,usenatbib]{mnras}
\usepackage{latexsym,graphicx}
\usepackage{amsmath}

\newcommand\msun{\, \rm M_\odot}
\newcommand\pc{{\, \rm pc}}

\def\gtsima{$\; \buildrel > \over \sim \;$}
\def\ltsima{$\; \buildrel < \over \sim \;$}
\def\gtrsim{\lower.5ex\hbox{\gtsima}}
\def\lesssim{\lower.5ex\hbox{\ltsima}}

\def\mbin{M_{\rm bin}}
\def\jbin{J_{\rm bin}}
\def\rlc{r_{\rm lc}}
\def\jlc{J_{\rm lc}}
\def\rj{\mathcal{R}_J}
\def\rjz{\mathcal{R}_{Jz}}

\def\Gadget{{\sc Gadget2}\,}
\def\Griffin{{\sc Griffin}\,}
\def\phiGRAPE{{$\phi$\sc GRAPE}\,}
\def\smile{{\sc SMILE}}
\def\sapporo{{\sc Sapporo}}

\def\apjl{ApJL}%
\def\apjs{ApJS}%
\def\aap{A\&A}%
\title[Collisionless losscone refilling] {Collisionless loss-cone refilling: there
  is no final parsec problem}

\author[Gualandris et al.]  
  {Alessia Gualandris$^{1}$\thanks{E-mail:
      a.gualandris@surrey.ac.uk}, Justin I. Read$^{1}$, Walter Dehnen$^{2}$ and Elisa Bortolas$^3$\\
  $^{1}$Department of Physics, Faculty of Engineering and Physical Sciences, University of Surrey, Guildford, GU2 7XH, United Kingdom\\
  $^{2}$Department of Physics and Astronomy, University of Leicester, Leicester, LE1 7RH, United Kingdom\\
  $^{3}$INAF-Osservatorio Astronomico di Padova, Vicolo dell'Osservatorio 5, I-35122, Padova, Italy\\
}

\begin{document}

\date{}

\maketitle

\begin{abstract}
  Coalescing massive black hole binaries, formed during galaxy
  mergers, are expected to be a primary source of low frequency
  gravitational waves. Yet in isolated gas-free spherical stellar
  systems, the hardening of the binary stalls at parsec-scale
  separations owing to the inefficiency of relaxation-driven loss-cone
  refilling. Repopulation via collisionless orbit diffusion in
  triaxial systems is more efficient, but published simulation results
  are contradictory. While sustained hardening has been reported in
  simulations of galaxy mergers with $N\sim 10^6$ stars and in early
  simulations of rotating models, in isolated non-rotating triaxial
  models the hardening rate continues to fall with increasing $N$, a
  signature of spurious two-body relaxation.

 We present a novel approach for studying loss cone repopulation in
 galactic nuclei. Since loss cone repopulation in triaxial systems
 owes to orbit diffusion, it is a purely collisionless phenomenon and
 can be studied with an approximated force calculation technique,
 provided the force errors are well behaved and sufficiently small. We
 achieve this using an accurate fast multipole method and define a
 proxy for the hardening rate that depends only on stellar angular
 momenta. We find that the loss cone is efficiently replenished even
 in very mildly triaxial models (with axis ratios 1 : 0.9 : 0.8). Such
 triaxiality is unavoidable following galactic mergers and can drive
 binaries into the gravitational wave regime. We conclude that there
 is no `final parsec problem'.
\end{abstract}


\section{Introduction}
Supermassive black hole binaries (BHBs) are expected to form
efficiently through cosmic time as a result of galaxy mergers, if the
original galaxies contain a central massive black hole (MBH)
\citep{BBR1980}. In the early stages of the merger, the MBHs are
dragged towards each other by dynamical friction, until they form a
binary system. The binary continues to evolve and harden by
gravitational interactions with stars that come within a few binary
separations. As a result of such encounters, stars subtract energy and
angular momentum from the binary, and they are ejected to larger
distances. $N$-body simulations of galaxy mergers
\citep[e.g.][]{QH1997, MM2001, GM2012} show a decrease in the central
stellar density and the formation of a core. This process, often
called {\it core scouring}, may be responsible for the observation of
cores in massive elliptical galaxies
\citep{Lauer1985,Lauer1995,faber1997,Graham2004,lauer2005,ferrarese2006,Graham2013}. If
a sufficient supply of stars is available in the merger remnant to
interact with the binary, hardening continues down to separations
where emission of gravitational waves becomes important. The evolution
then proceeds rapidly to inspiral and black hole coalescence.  The
recent detections of gravitational wave signals (GW150914 and
GW151226) by Advanced Ligo \citep{GW150914, GW151226} from the merger
of two stellar mass black holes proves very strongly that black holes
exist, they form binaries, and merge within a Hubble time due to
emission of gravitational waves. The same is expected for MBHs, and
several missions and detectors are being planned in the hope to detect
gravitational waves from BHBs. In the context of BHBs, The Pulsar
Timing Array (PTA) \citep[e.g.][]{Babak2016}, which is already
operational, is the most suitable to detect gravitational waves from
the most massive BHBs, while the $e$LISA space-based interferometer
\citep{Barausse2015} will be sensitive to the frequencies typical of
lower mass BHBs, like the Milky Way's MBH. Detection of BHBs in the
gravitational waves window would provide crucial information on the
masses, spins, orientations and even distances of the two black
holes. It would also provide an exciting new cosmological probe, given that
 BHBs can in principle be seen right back to the
beginning of the Universe \citep{hogan2009}.

All efforts to detect gravitational waves from BHBs are based on the
assumption that the binaries do in fact coalesce efficiently, i.e. a
significant fraction have a timescale for emission of gravitational
radiation shorter than a Hubble time.  This assumption seems to be
corroborated by observations: despite strong efforts to detect BHBs
with a variety of techniques, only a handful of candidates exist, and
for most of these systems alternative explanations have been put
forward \citep[for a review see e.g.][]{DSD2012}. It appears that BHBs
find a way to coalescence, but the mechanism for this remains to be
understood.  While gas may play a role in bringing BHBs into the
gravitational wave regime, there is evidence for little or no gas in
massive elliptical galaxies, and works by \citet{cuadra2009} and
\citet{lodato2009} show that gas discs are only efficient at driving
mergers of BHBs for low mass MBHs. This suggests that if BHBs merge,
in many cases they must do so in the absence of gas. In this paper, we
consider only pure stellar systems devoid of gas.

Theoretically, the ultimate fate of BHBs depends on the
supply of stars to the binary's loss cone, i.e. the region in phase
space characterised by low enough angular momentum to ensure
interaction with the binary. If the reservoir of stars on intersecting
orbits is depleted and the loss cone cannot be maintained sufficiently
populated, the binary's evolution will dramatically slow down or even
stall \citep{BBR1980}.

There are two distinct classes of physical mechanisms for loss cone
refilling in the absence of gas: collisional processes and collisionless processes.  In
idealised spherical galaxies, the evolution of BHBs slows down after
all stars initially on loss cone orbits are ejected, which occurs on a
dynamical timescale. Repopulation of the loss cone in such models is
due solely to two body relaxation, i.e. a collisional process. This
operates on the relaxation timescale, which for most galaxies is much
longer than the Hubble time, and implies that the binary is in the
empty loss cone regime at all times \citep{MM2001, MM2003}.  In $N$-body
simulations of isolated spherical models, this regime can be
identified by the $N$-dependence of the binary hardening rate, since
the relaxation time scales approximately as $N/{\rm log}N$
\citep{MF2004,BMS2005,merritt2007}. The earliest simulations, however,
did not observe this scaling due to the low $N$ used which implied a
very short relaxation time and therefore a full loss cone
\citep[e.g.][]{MM2001}.  Collisional processes are rather inefficient
at repopulating the loss cone in real galaxies, and are not sufficient
to lead BHBs to coalescence in a Hubble time. However, they need to be
treated very carefully since they introduce a spurious population of
loss cone orbits in any $N$-body simulation where $N$ is smaller than
the true number of stars.

In non-spherical galaxies, refilling of the loss cone can proceed due
to collisionless ``diffusion'' in angular momentum. Torques from a
non-spherical potential cause stellar angular momenta to change on a
timescale which is much shorter than the relaxation time, though
typically longer than radial orbital periods \citep{Merrittbook,pontzen2015}. In
axisymmetric nuclei, stellar orbits conserve only energy and the
component $J_z$ of angular momentum parallel to the symmetry
axis. While the total angular momentum $J$ is not conserved,
conservation of $J_z$ implies that there is a lower limit to how small
$J$ can become due to diffusion.  In this sense, orbits in
axisymmetric potentials are not truly centrophilic, but they reach very
small radii and contribute to loss cone refilling. There are two
families of orbits in flattened potentials: the tube orbits, regular
orbits for which $J$ is approximately conserved, and saucer orbits,
for which $J$ can become very small. The fraction of stars on saucer
orbits depends on the degree of flattening in the system.  In triaxial
nuclei, $J$ is not conserved and truly centrophilic orbits are
possible. In addition to two families of tube orbits and saucer
orbits, there exists a new family of orbits, called pyramids, which
can achieve arbitrarily low values of $J$ \citep{MV2011}. All stars on
pyramid orbits eventually reach the centre, though the timescale
can vary.  While the fraction of saucer orbits in axisymmetric models
is expected to be small, the fraction of pyramid orbits in triaxial
models can be very high \citep{PM2004}.  This suggests that
collisionless loss cone refilling may be efficient at driving BHBs to
coalescence \citep[e.g.][]{NS1983, PM2004}.

Supporting evidence for collisionless loss cone refilling appeared
with the first simulations of merging galaxies hosting central MBHs,
in which the merger was followed from early times
\citep{preto2011,khan2011,GM2012,khan2012} as well as in
cosmologically motivated mergers \citep{khan2016}. In these
simulations, the hardening rate of the binary is found to be to
largely independent of $N$, and the merger remnant shows significant
triaxiality, at least in the central regions. The supply of stars in
these cases is sufficiently high to ensure MBH coalescence in much
less than a Hubble time.  However, the complexity of galaxy merger
simulations leaves open the possibility that a different process (that
could be physical or a numerical error) is responsible for the
sustained binary. If torques from a non-spherical background are
responsible for efficient loss cone refilling, the same behaviour
should be observed for binaries placed in isolated non-spherical
models.  The first $N$-body simulations of BHBs in triaxial galaxy
models were performed by \citet{BMSB2006}, who considered flattened
models with net rotation and $N \leq 10^6$. In this case, an equal
mass BHB is found not to stall but to sustain an hardening sufficient
to lead to coalescence.  On the other hand, \citet{VAM2014} follow the
evolution of BHBs in spherical, axisymmetric and triaxial non-rotating
isolated models, using direct $N$-body simulations with particle
numbers up to one million. They find an $N$ dependence of the
hardening rate in all models, with only a mild flattening at the
largest $N$ for the non-spherical models. Hardening rates are higher
in non-spherical models than in spherical ones, but always lower than
the full loss cone rate. They conclude that collisional effects still
contribute significantly to the loss cone repopulation at these $N$,
and prevent a reliable extrapolation to real galaxies.  $N$-body
simulations of collisional stellar systems are very expensive,
especially when regularisation and/or very small softening are
employed to model the evolution of the BHB accurately.  A possible
solution to limit the effects of collisional repopulation is to adopt
a collisionless numerical method.  \citet{VAM2015} follow the
evolution of BHBs in isolated models with a Monte Carlo code able to
suppress the effects of two-body relaxation. In this case, they find
that hardening rates tend to become $N$-independent in triaxial models
in the limit of large $N$ ($N \gtrsim 5\times10^6$). Rates are always
lower than the full loss cone rate, but in triaxial models they are
sufficient to drive BHBs to coalescence in less than a Hubble
time. Axisymmetric models, however, have hardening times that are too
long to be of interest. This is in contradiction with the results of
\citet{khan2013} for axisymmetric models, and the reason for the
discrepancy remains unknown, though likely of numerical origin.

In this study, we present an alternative approach to model the loss
cone refilling of BHBs. By means of direct summation simulations of
galaxy mergers, we show that a good proxy for the binary hardening
rate is given by the integrated number of stars with angular momentum
smaller than the loss cone angular momentum of the binary. We estimate
this at the hard-binary separation, counting each star only once in
the time interval of interest. This is because we expect stars to be
scattered out of the loss cone following an interaction. We then
consider different isolated galaxy models, spherical and flattened,
and follow their evolution with both a direct summation code and a
tailored fast multiple method. We show that the hardening rate is a
strong function of particle number $N$ in spherical models, with a
scaling consistent with expectations for collisional processes. On the
other hand, the hardening rate becomes independent of N for $N \gtrsim
5\times 10^6$, in triaxial models.  The behaviour of axisymmetric
models is intermediate between that of spherical and triaxial models,
with a slow but significant dependence on $N$.  We reach values of $N$
equal of 2 million particles with direct summation and 64 million
particles with the fast multiple method. Efficient loss cone refilling
is seen even in mildly triaxial models (with axis ratios
$1:0.9:0.8$). Such triaxiality is unavoidable following galaxy mergers
and drives binaries into the gravitational waves regime. We conclude
therefore that there is no `final parsec problem' in the evolution of
BHBs even in gas-free systems due to the efficiency of collisionless
loss cone refilling in triaxial potentials.

\section{The hardening rate of black hole binaries}
\label{sec:theory}
In this section, we present a semi-analytic model for binary
hardening. We will compare this with numerical simulations in section
\ref{sec:sims}.  Consider a binary system of two MBHs with masses
$M_1$ and $M_2$ and semi-major axis $a$. The total mass of the binary
is $\mbin = M_1+M_2$ and the binding energy is
\begin{equation}
E = \frac{GM_1M_2}{2a} = \frac{G\mu \mbin}{2a}
\end{equation}
where $\mu = M_1M_2/\mbin$ is the reduced mass.  The time-dependent hardening rate of the binary is defined as
\begin{equation}  
\label{eq:s}
  s \equiv \frac{d}{dt} \left(\frac{1}{a}\right) . 
\end{equation}

We define a binary to
be ``hard'' if the binding energy per unit mass exceeds $\sigma^2$,
where $\sigma$ is the stellar velocity dispersion. The hard binary separation can be defined as
\begin{equation}
a_h = \frac{G\mu}{4\sigma^2} = \frac{M_2}{\mbin} \frac{r_h}{4}
\end{equation}
where $r_h = GM_1/\sigma^2$ is the influence radius of the most
massive MBH.
Stars of angular momentum $J$ interact with the binary if $J \leq \jlc$, the
loss cone angular momentum
\begin{equation}
\jlc = {\mathcal{K}} a \sqrt{2 [E-\Phi(\mathcal{K} a)]} \approx
\sqrt{2G\mbin \mathcal{K}a}.
\end{equation}
Here $\mathcal{K}$ is a dimensionless constant of order unity, and in
the following we adopt $\mathcal{K}=1$.

Following the classical analysis by \citet{Hills1983} we define a
dimensionless coefficient describing the energy exchange in a single
encounter between the binary and a single star of mass $m$ 
\begin{equation}  
C = \frac{\mbin}{2m} \frac{\Delta E}{E},
\end{equation}  
where $\Delta E = -(G\mbin/2) \Delta(1/a)$ is the change in binding energy.
In the hard binary limit $a \ll a_h$, the outcome of the interaction
depends solely on the ratio $\chi = J/\jbin$, where $\jbin = \sqrt{2G\mbin a}$ is
the angular momentum of the binary, assumed circular for simplicity.

The change in binary hardness in one encounter is simply:
\begin{equation} 
\Delta \left(\frac{1}{a}\right) = \frac{2m}{\mbin a} C(\chi)\,.
\end{equation}

The hardening rate of a  binary in a fixed density background with uniform density
$\rho$ and uniform velocity $v$ is the \citep{VAM2014}
\begin{equation} 
s = \frac{\rho v} {m} \int_0^{\infty} \Delta \left(\frac{1}{a}\right)
2\pi b db = \frac{G \rho}{v}  \int_0^{\infty} 8\pi C(\chi) \chi d\chi.
\end{equation} 
Here we adopt the expression for $C(\chi)$ derived by \citet{VAM2014}
for the case of an equal mass circular binary based on the dependence
found by \citet{sesana2006}.
If we identify 
\begin{equation}
H \equiv \int_0^{\infty} 8\pi C(\chi) \chi d\chi, 
\end{equation}
this is of the form of the expression derived by \citet{Quinlan1996}
\begin{equation} 
s_H = \frac{G\rho}{v} H
\end{equation} 
with $H$  a dimensionless hardening coefficient, which reaches an
approximately constant value $H_0\approx 16$ in the hard binary limit $a
\ll a_h$. This result implies that a hard binary hardens at a constant rate in a
constant density background.
Stellar densities do not remain constant during galaxy mergers due to
the slingshot ejection of loss-cone stars. A constant or slowly
declining hardening rate therefore implies an efficient repopulation
of the loss-cone in merger simulations \citep{preto2011,khan2011, GM2012}.

The hardening rate can be computed in an approximate way directly from
the $N$-body data as the summation of the individual contributions of
all stars. In this model, we have
\begin{equation} 
\label{eq:hardnb}
s_{NB} = \sum_{i=1}^N \frac{2m_i}{\mbin P_i} \frac{C(\chi)}{a},
\end{equation} 
where $m_i$ is the mass of the star and $P_i$ its radial orbital
period.  In principle, the summation extends to all stars in the
system. However, the function $C(\chi)$ quickly falls to zero for
$\chi >1$, with no contribution from stars with $\chi \gtrsim 2$
\citep[fig.1]{sesana2006}.

In the next section, we show that the hardening rates computed from
the $N$-body data are in good agreement with those measured from the
binary's semi-major axis in merger simulations. We also show that
restricting the summation in equation~\ref{eq:hardnb} to stars with $\chi
\leq 1$ still provides a good match while significantly reducing
computational cost. We then introduce a proxy for the hardening rate of
the binary which depends only on the integrated number of stars with
$\chi \leq 1$ over the simulation time and show that, although
approximate, this allows us to study loss cone refilling in the
collisionless regime.

\section{Numerical simulations}
\label{sec:sims}

\subsection{Simulations of galaxy mergers}
\label{sec:mergers}

First we perform collisional direct summation $N$-body simulations of mergers of galaxies
harbouring central massive black holes and measure the hardening rate of the
black hole binary.  For these simulations, we make use of the HiGPUs
\citep{HiGPU2013} GPU parallel software. We consider spherical
galaxy models following Dehnen's spherical density profile
\citep{Dehnen1993, saha1993}
\begin{equation}
  \label{eq:dehnen}
  \rho (r) = \frac{\left(3-\gamma\right) M}{4\pi} \frac{a}{a^{\gamma} \left(r+a\right)^{4-\gamma}}
\end{equation}
where $M$ is the total mass of the galaxy, $a$ the scale radius and
$\gamma$ the asymptotic inner slope.  For $\gamma=1$
equation~\ref{eq:dehnen} gives the Hernquist model
\citep{Hern1990}. We adopt standard $N$-body units $M = G = a = 1$.

We perform one set of simulations of equal mass galaxies with
$\gamma=1$ and different $N$ from $2^{15} \sim $ 32k to $2^{20} \sim 1$ million. Each model
contains a supermassive black hole of mass $0.005$, in units of the
galaxy mass.  The galaxies are placed on a bound elliptical orbit with
eccentricity $e=0.4$ and at an initial distance $D = 5$.  The
softening length is set to $\epsilon = 10^{-4}$. We follow
the evolution of the binary for a time $t_{\rm end} = 200$. In these
units, the formation of a hard binary occurs at $t \sim 50$.

\begin{figure}
  \begin{center}
    \includegraphics[height=8cm, angle=270]{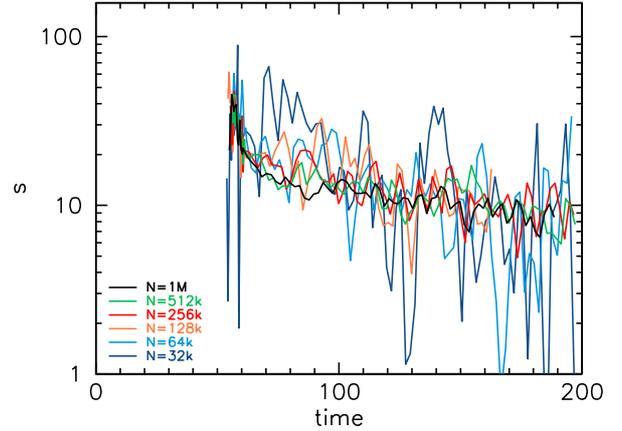}
  \end{center}
  \caption{Time dependence of the hardening rate in collisional
    simulations of galaxy mergers with different particle number $N$,
    up to one million.}
  \label{fig:stime}
\end{figure}

\begin{figure}
  \begin{center}
    \includegraphics[height=8cm, angle=270]{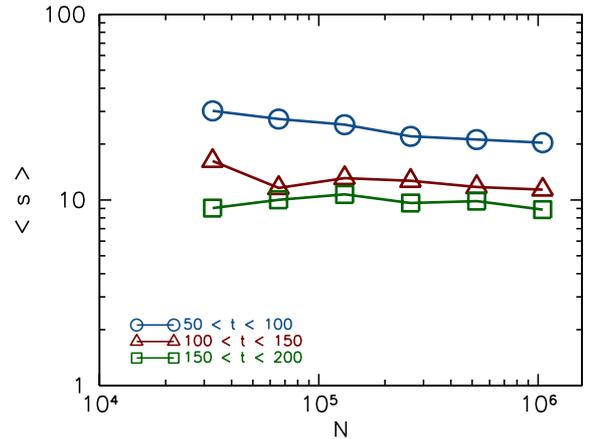}
  \end{center}
  \caption{Binary hardening rate in direct collisional simulations of
    galaxy mergers with different particle number $N$, averaged over
    different time intervals: $50 \leq t \leq 100$ (circles), $100
    \leq t \leq 150$ (triangles) and $150 \leq t \leq 200$ (squares).}
  \label{fig:sN}
\end{figure}

We monitor the evolution of the binary's semi-major axis $a$ and
compute the hardening rate by fitting a straight line to $a^{-1}(t)$
in small time intervals, from the time of binary formation to the end
of the integration. The result is shown in Fig.~\ref{fig:stime} for
models with different particle numbers.  Naturally, low-$N$ models
show considerable noise but we include them for completeness. As the
binary shrinks due to gravitational slingshot ejections of interacting
stars, the loss cone angular momentum $\jlc$ decreases and the hardening rate
decreases slowly with time.
However, it does so for all considered choices of $N$, and shows no significant
$N$-dependence.
We also consider the hardening rate averaged over a given
time-interval: specifically over $50 < t<100$, $100<t<150$ and
$150<t<200$.  The first is meant to capture the phase of quick
hardening after the binary becomes formally bound and then
reaches the hard-binary separation, while the second and third
represent the late phases of hardening, once all the stars initially
in the loss cone have been ejected and hardening is due to refilling of
the loss cone.  The averaged hardening
rates are shown in Fig.~\ref{fig:sN} as a function of particle number.
Despite the noise that plagues low-$N$ simulations, there is
essentially no dependence of the hardening rate on particle number.
This is in agreement with the results of earlier simulations of galaxy
mergers \citep{preto2011,khan2011, GM2012}.  We find a similar result
for a second set of simulations of mergers of equal mass galaxies with
a $\gamma=1.5$ density profile.

We also compute the hardening rate in the merger simulations directly
from the $N$-body data, according to equation~\ref{eq:hardnb}, both
summing up the contributions of all stars and considering only
stars with $\chi \leq 1$. For the computation of $C(\chi)$ we adopt
the fitting function given in \citet[equation 5]{VAM2014} which well
approximates the results of \citet{sesana2006} in the case of a
circular, equal mass binary. The radial orbital period of each star is
computed by solving numerically the integral
\begin{equation}
\label{eq:trad}
P_i = 2 \int_{r_p} ^{r_a} \frac{dr}{\sqrt{2 [E-\Phi(r)] - J^2/r^2}},
\end{equation}
where $E$ and $J$ are, respectively, the energy and angular momentum
of the orbit, $\Phi(r)$ is the gravitational potential and the pericentre $r_p$ and apocentre $r_a$ are
computed by solving numerically the equation for the turning points in
a spherical potential
\begin{equation}
\label{eq:trad}
\frac{1}{r^2} + \frac{2[\Phi(r) - E]}{J^2} = 0.
\end{equation}

\begin{figure}
  \begin{center}
    \includegraphics[height=8cm, angle=270]{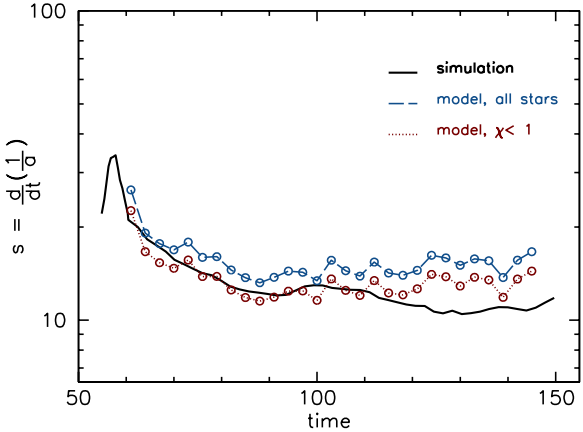}
  \end{center}
  \caption{Binary hardening rate versus time measured in a galaxy
    merger with $N=10^6$ (solid line) and estimated from the stellar
    angular momenta according to equation~\ref{eq:hardnb} including all
    stars (dashed line) and only stars with $\chi \leq 1$ (dotted line).
}
  \label{fig:hardnb}
\end{figure}

The hardening rates computed for the $N=2^{20} \sim 10^6$ merger simulation are shown in
Fig.~\ref{fig:hardnb}, and compared with the rate measured from the
binary's semi-major axis according to the definition given in
equation~\ref{eq:s}. The agreement is quite good, also in the case when the
contribution from stars with $\chi >1$ is neglected.

We now look for a simple proxy for the binary hardening rate that
depends only on the angular momenta of the stars. Given that, at any
time, the hardening of the binary is due to stars with angular
momentum smaller than the loss cone angular momentum $J \le \jlc$, we
consider as proxy for the hardening rate the parameter
\begin{equation}
\label{eq:pp}
\rj \equiv \frac{N_J}{N} = \frac{N (J \le \jlc)}{N}
\end{equation}
where $N_J$ is the total number of stars with initial angular momentum
larger than the loss cone angular momentum at the hard binary separation
$J>\jlc (a_h)$ but which reach $J< \jlc$ at some point in the
simulation. In particular, stars moving in and out of the loss-cone
are counted only once, to account for the fact that they would be
ejected after a close encounter with the BHB.
We will refer to $\rj$ as the refilling parameter.

\begin{figure}
  \begin{center}
    \includegraphics[height=8cm, angle=270]{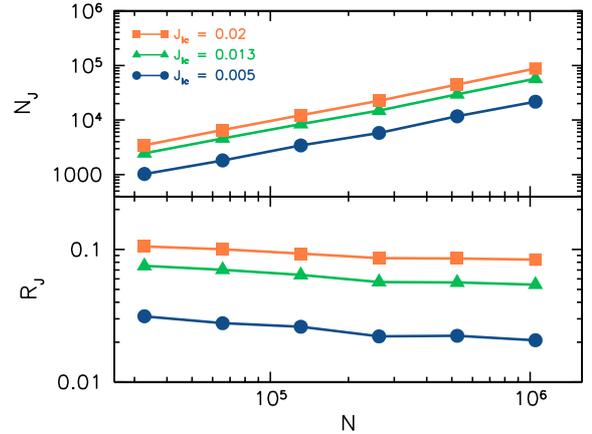}
  \end{center}
  \caption{Top: total number of stars with angular momentum $J \leq \jlc$, representative of the
    loss cone angular momentum at the hard binary separation for
    binaries with different total mass. $N_J$ was computed in the merger simulations with
    initial $\gamma=1.0$ profile. Each star satisfying the
    condition is counted only once and stars satisfying the condition
    at the start of the simulations are discounted. Bottom: refilling
    parameter, as defined in equation~\ref{eq:pp}, in the merger simulations.}
  \label{fig:mergerpop}
\end{figure}

We compute the loss cone refilling parameter in the merger
simulations, where we assumed three different values for the loss cone
angular momentum, representative of binaries with different total mass
in units of the galaxy mass. The results, shown in
Fig.~\ref{fig:mergerpop}, indicate that while $N_J$ is a strong
function of $N$  for all considered values of $\jlc$ (see top panel),
the refilling parameter $\rj$ is effectively independent of $N$ (see
bottom panel).
The refilling parameter is therefore a reliable proxy for the
binary hardening rate: if $\rj$ is independent of $N$, so is the
hardening rate.

\subsection{Simulations of isolated models}
\label{sec:isolated}
The refilling parameter depends only on the number of stars with
angular momentum smaller than a given critical value, i.e. only on the
fraction of stars able to interact with the binary during the
hardening phase. This, in turn, depends on the level of non-sphericity
of the gravitational potential.  Therefore, it is not necessary to
perform simulations of galaxy mergers with massive black holes in
order to model collisionless loss cone refilling.  Such simulations are required
for a precise determination of the hardening rate and the coalescence
timescale, but not to ascertain the fate of BHBs in the broader
context of the final parsec problem.  Simulations including BHBs are
computationally very expensive because of two main reasons: (i) direct
summation is usually implemented for the computation of 
gravitational forces (ii) a small gravitational softening is adopted
to follow the evolution of the binary to very small separations, well
below the hard-binary limit.

Here we take a different approach in which we simulate isolated galaxy
models without massive black holes and estimate the collisionless loss
cone refilling rate by measuring the refilling parameter in models
with increasing $N$. We adopt both a direct summation technique, with
particle numbers up to $2^{21} \sim$ 2 millions, and a tailored fast
multiple method, with particle numbers up to $2^{26}\sim$ 64 million.
For the former set of simulations we use the {\phiGRAPE} code
\citep{harfst2007}, adapted to run on a GPU cluster by means of the
{\sapporo} library \citep{sapporo2009}.  For the latter set of
simulations we adopt the collisionless $N$-body code {\Griffin}, which
employs the fast multipole method (FMM). To allow for softened
gravity, the FMM is implemented in Cartesian coordinates, but
otherwise is very similar to the method reported in
\citep{Dehnen2014}, in particular the computational effort at a given
accuracy is minimised while maintaining a well behaved distribution of
errors. In this way, the code can be made as accurate as direct
summation with a $\mathcal{O}(N)$ scaling. The simulations reported
here use multipole expansion order $p=5$ and a relative force error of
$5\times10^{-5}$ for each particle. Initially, we attempted to use the
public $N$-body code {\Gadget}, but could not obtain converged results
(see Appendix A), most likely owing to insufficient force accuracy of
the tree method.

Very large particle numbers are required to eliminate collisional
effects and measure collisionless repopulation of the loss cone, as
these occur on a relaxation time scale, which scales as $N/{\rm ln}
N$.  However, we also perform direct summation simulations to validate
the results of the fast multiple method code and to compare with
existing results.

We consider galaxy models following a generalisation of the
\citet{Hern1990} profile
\begin{equation}
\rho(r) = \frac{M}{2\pi\, abc} \frac{1}{r \left(1+r\right)^3}
\end{equation}
to allow for non-sphericity, where $a$, $b$ $c$ represent the axes of
a triaxial ellipsoid. Units are such that $M=1$, where $M$ is the total mass of the
galaxy, and $abc=1$.

The parameters of all the models are listed in Table \ref{tab:par},
including spherical models, axisymmetric models with axis ratios
1:1:0.8 and two sets of triaxial models, moderately triaxial with
1:0.9:0.8 and triaxial with 1:0.8:0.6. 

\begin{table}
  \begin{tabular}{llllll}
    \hline
    Model & Axis ratios & $M_{\rm bin} / M$ & $r_m$ & $a_h$ & $J_{\rm LC}$\\
\hline
S & 1:1:1 & 0.001 & 0.13 & 0.008 & 0.004\\
S & 1:1:1 & 0.005 & 0.22 & 0.014 & 0.012\\
S & 1:1:1 & 0.01 & 0.28 & 0.018 & 0.018\\
\hline
A & 1:1:0.8 & 0.001  & 0.13 & 0.008 & 0.004\\
A  & 1:1:0.8 & 0.005 & 0.22 & 0.014 & 0.012\\
A  & 1:1:0.8 & 0.01 & 0.28 & 0.017 & 0.019\\
 \hline
T1 & 1:0.9:0.8 & 0.001  & 0.25& 0.015 & 0.0055\\
T1 & 1:0.9:0.8 & 0.005 & 0.41 & 0.025 & 0.016\\
T1 & 1:0.9:0.8 & 0.01 & 0.50 & 0.032 & 0.025\\
\hline
T2 & 1:0.8:0.6 & 0.001 & 0.26 & 0.016 & 0.0057\\
T2 & 1:0.8:0.6 & 0.005 & 0.44 & 0.027 & 0.016\\
T2 & 1:0.8:0.6 & 0.01 & 0.55 & 0.034 & 0.026\\
\hline\hline 
\end{tabular}
\caption{Parameters used to estimate the loss cone angular momentum
  in spherical (S), axisymmetric (A) and triaxial (T) models. The columns
  indicate, respectively, the model name, the axis ratio, the mass of
  the hypothetical black hole binary in units of the total galaxy
  mass, the influence radius of the black hole
  binary, the hard-binary separation and the resulting loss cone angular
  momentum.}
  \label{tab:par}
\end{table}
All models were generated with the {\smile} software
\citep{vasiliev2013}, a recent implementation of the Schwarzschild
orbital superposition method, using $2\times 10^5$ orbits.  

We adopt an $N$ dependent value of softening appropriate for the FMM
code of the order
\begin{equation}
\label{eq:soft}
\epsilon \approx \left(\frac{2\pi}{N}\right)^{\frac{1}{3}}
\end{equation}
and apply the same to the {\phiGRAPE} simulations. While this
choice is larger than the values generally employed in 
direct summation simulations, it is appropriate for the collisionless
process to be studied here. A more detailed description of the effect
of softening is given in section \ref{sec:softening}.

\begin{figure*}
  \begin{center}
    \includegraphics[height=8cm]{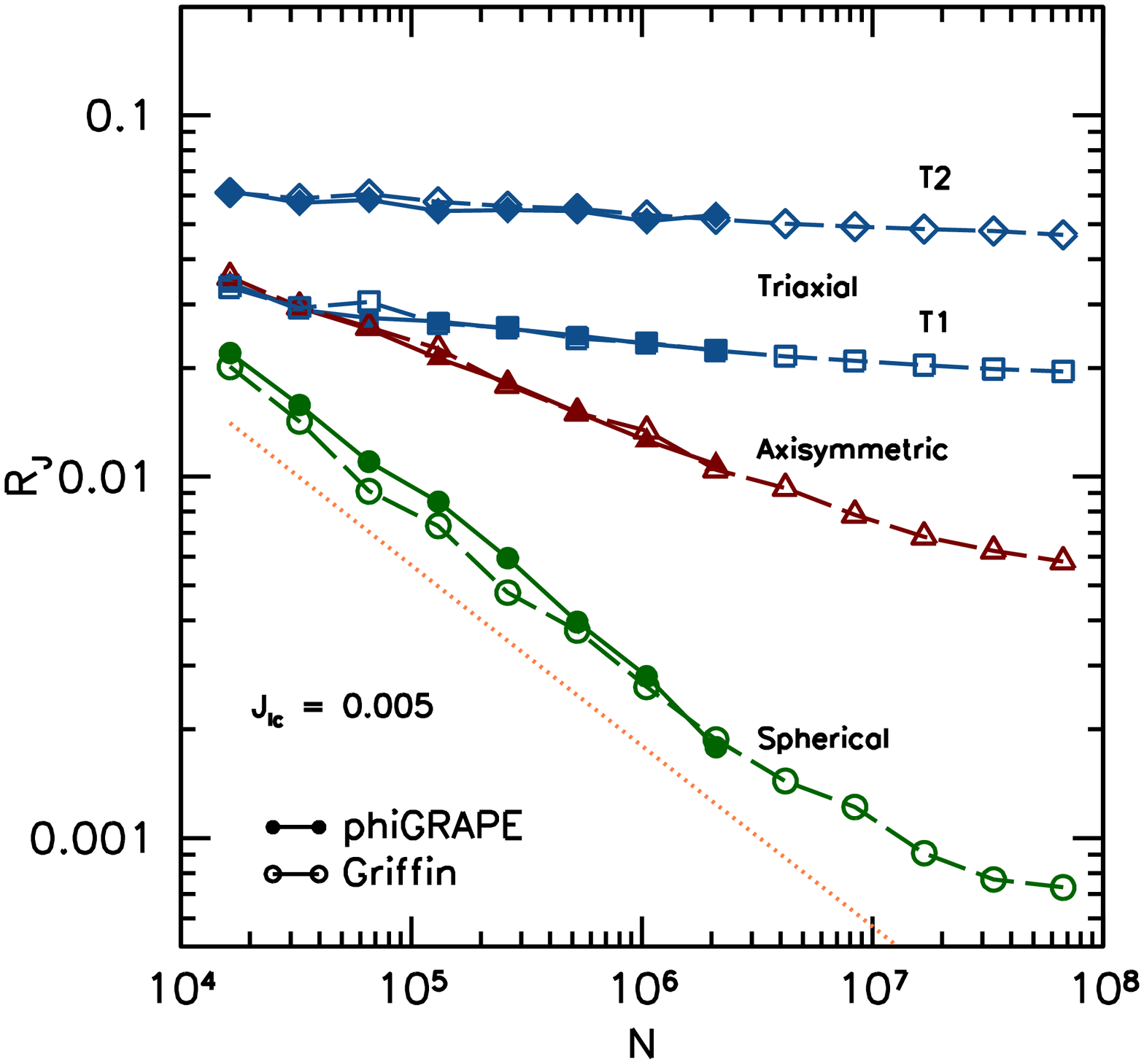}
    \includegraphics[height=8cm]{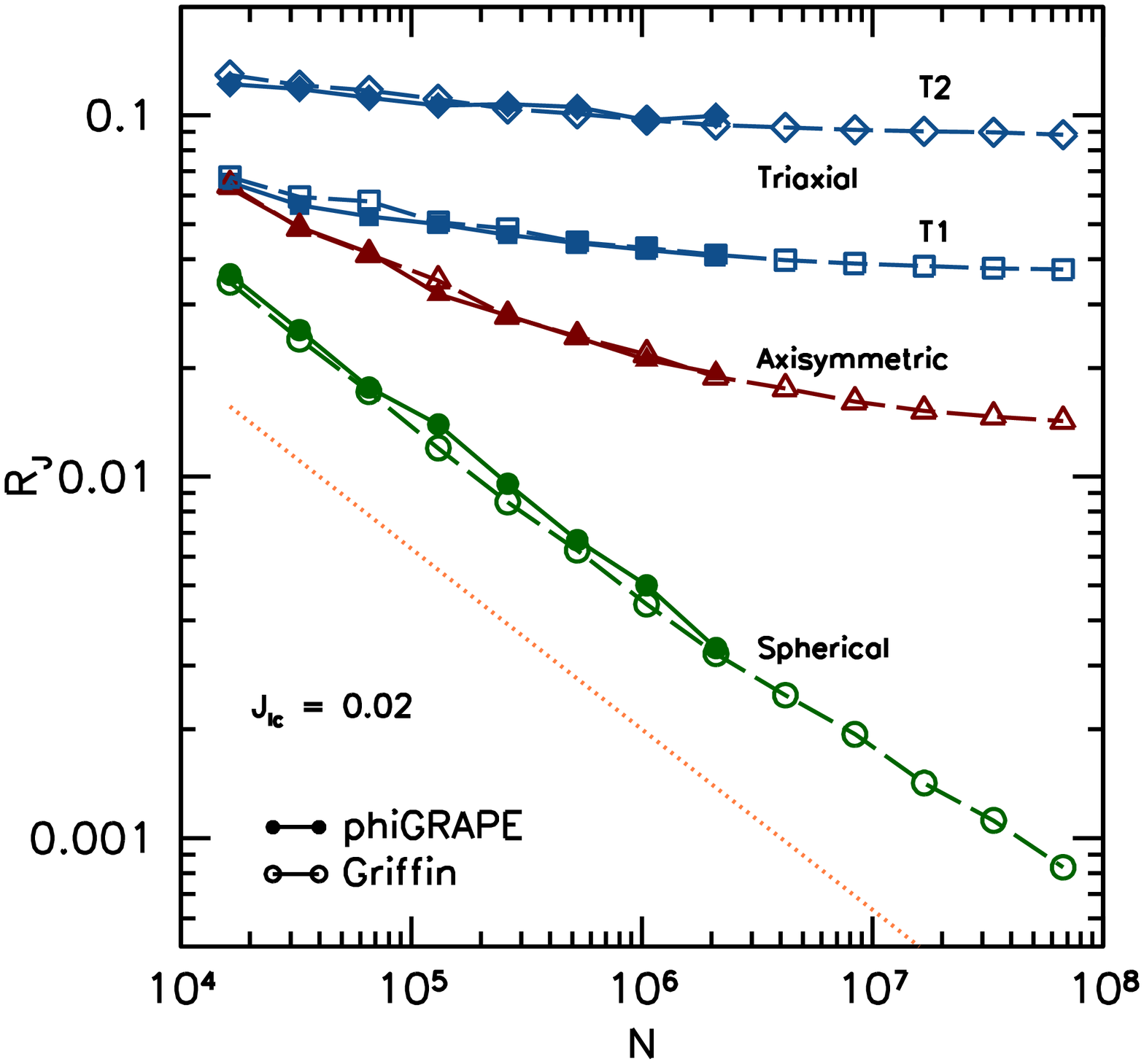}
  \end{center}
  \caption{Refilling parameter $\rj$ as a function of particle number
    for different models assuming a loss cone angular momentum
    $\jlc=0.005$ (left) and $\jlc = 0.02$ (right). In both panels the
    dotted line shows the $1/\sqrt{N}$ slope expected for collisional
    loss cone repopulation.}
  \label{fig:njj}
\end{figure*}

\begin{figure*}
  \begin{center}
    \includegraphics[height=8cm]{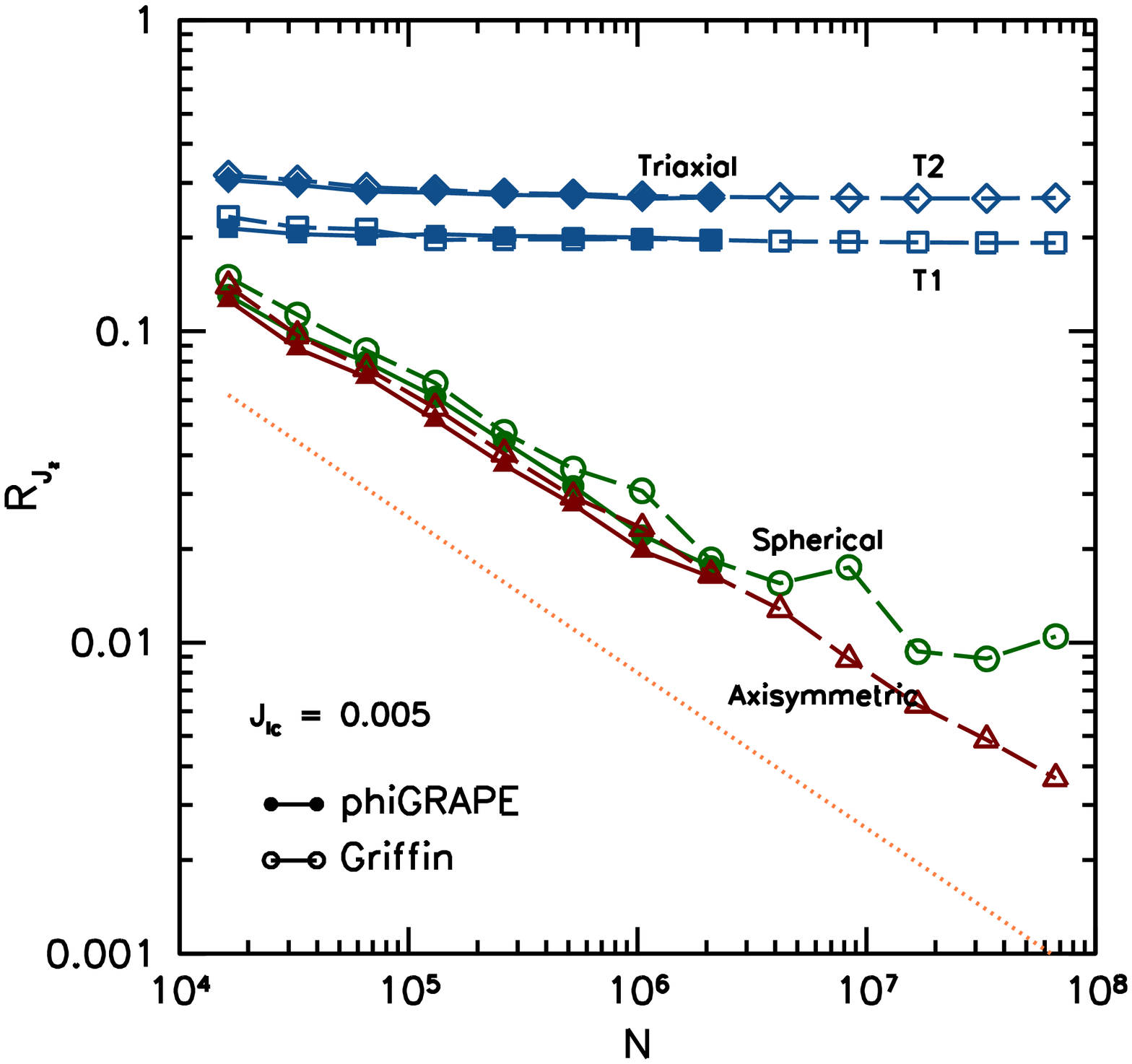}
    \includegraphics[height=8cm]{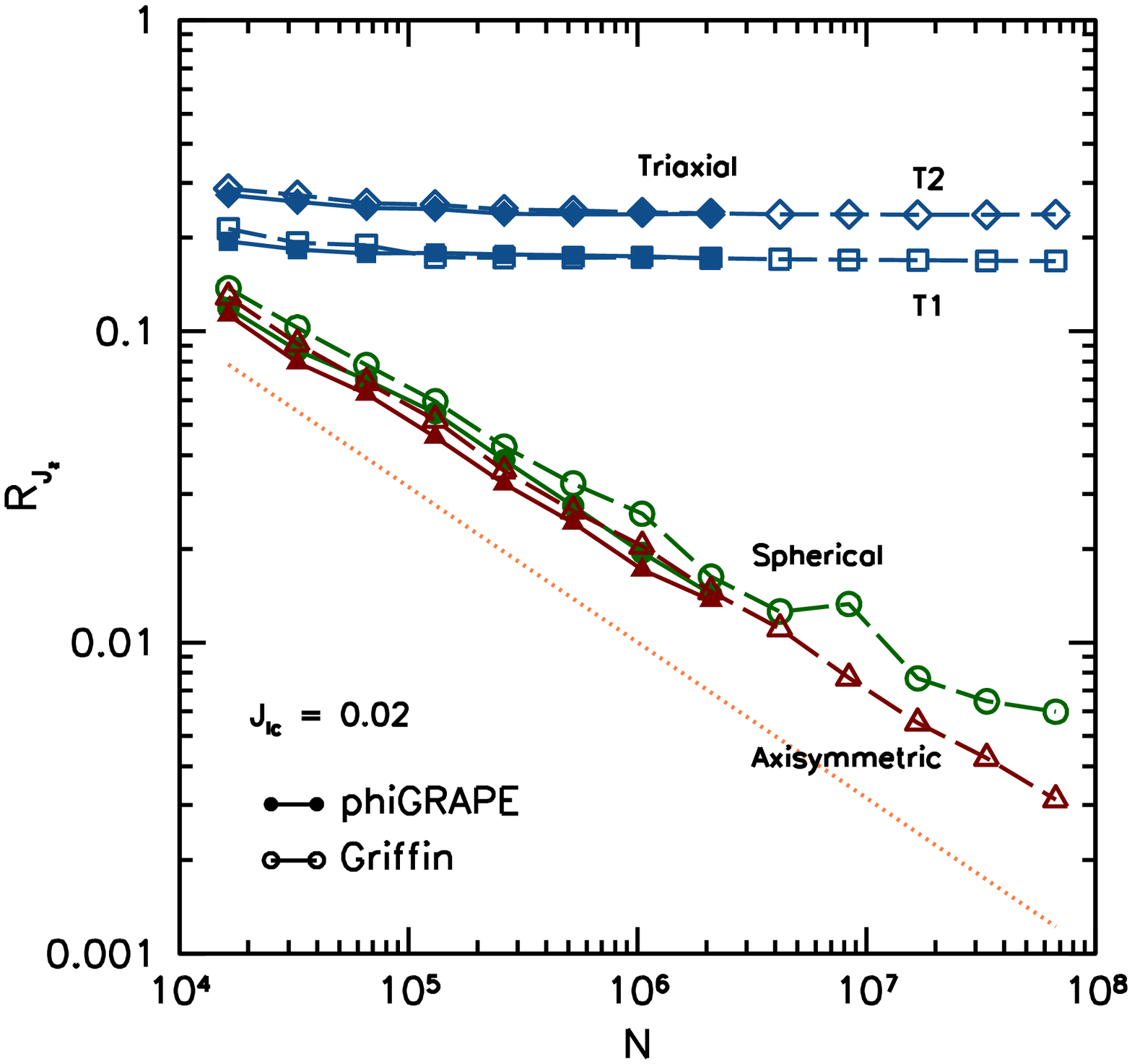}
  \end{center}
   \caption{Refilling parameter $\rjz$ as a function of particle number
    for different models assuming a loss cone angular momentum
    $\jlc=0.005$ (left) and $\jlc = 0.02$ (right). In both panels the
    dotted line shows the $1/\sqrt{N}$ slope expected for collisional
    loss cone repopulation.}
  \label{fig:njz}
\end{figure*}
Figure \ref{fig:njj} shows the refilling parameter $\rj$ as a
function of particle number for all the models and with both codes,
assuming a loss cone angular momentum $\jlc=0.005$ and $\jlc=0.02$.
These values correspond to the cases of binary to galaxy mass ratio of
0.001 and 0.01, respectively (see Table \ref{tab:par}).
We find that for spherical models $\rj$ decreases with $N$ as
$1/\sqrt{N}$, a signature that in these models collisional effects are
responsible for loss cone refilling. This holds true even at the
largest $N$ reached with \Griffin.  In triaxial models, instead, the
behaviour is dramatically different. Not only is $\rj$ much larger in
the triaxial models than in the spherical ones, there is essentially
little or no $N$ dependence, as expected if a collisionless process is
responsible for refilling the loss cone. The systematically larger
values of $\rj$ in the more triaxial models (T2) compared to the
moderately triaxial ones (T1) also supports the interpretation that the shape
of the potential is the key element in the refilling process.
Axisymmetric models show a behaviour which is intermediate between
that of the spherical and triaxial models, with a refilling parameter
which decreases as a function of $N$ as in spherical models, but much
more slowly. This can be interpreted as a consequence of the nature of
orbits in an axisymmetric potential, where the total angular momentum
$J$ of individual particles is not conserved but $J_Z$ is, and there
is therefore a lower limit to how small $J$ can become during the
evolution. In the case of the largest $\jlc$, there is evidence for
flattening at the largest $N$, but convergence is certainly not yet
reached in these models. The implication for BHBs in merger remnants
is therefore that flattening of the system may not be sufficient to
sustain orbital decay to the gravitational wave phase, but that a certain, even
moderate, degree of triaxiality may be needed to drive the binaries to
coalescence within a  Hubble time. This is in good agreement with the
findings of \citet{VAM2015}.
Figure \ref{fig:njz} shows the refilling parameter computed for just
the $J_z$ component of stellar angular momentum.  Because this
component is conserved for stars in an axisymmetric potential, $\rjz$
shows the same $N$ dependence in axisymmetric models as in spherical
models. On the other hand, there is no $N$ dependence in triaxial
models, as expected.

\begin{figure}
  \begin{center}
    \includegraphics[height=8cm]{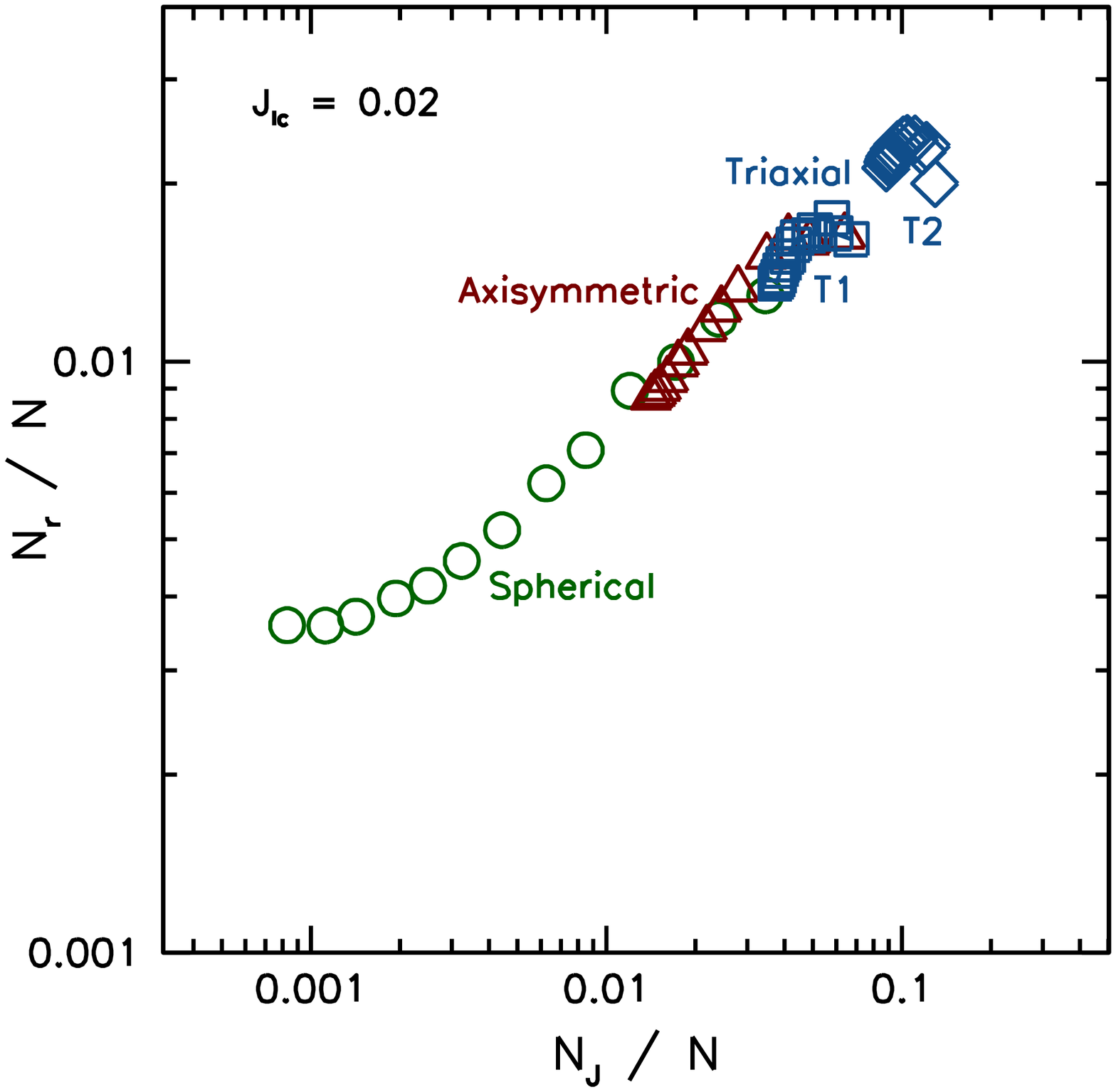}
  \end{center}
  \caption{Correlation between the number of stars $N_r$ within a
    given distance from the centre and the number of stars $N_J$ with
    angular momentum smaller than the loss cone angular momentum, for
    spherical (circles), axisymmetric (triangles) and triaxial
    (diamonds) models. All models were evolved with {\Griffin} and
    the refilling parameter is computed with $\jlc = 0.02$, while
    $N_r$ is computed for a distance $\rlc = 0.018$ from the centre,
    corresponding to the hard binary separation of a black hole binary
    with mass of 1\% of the galaxy mass.}
  \label{fig:njnr}
\end{figure}
Our computation of the refilling parameter as a proxy for the
hardening rate relies on the assumption that stars whose angular
momentum falls below the binary angular momentum populate the loss
cone and will eventually interact with the binary, contributing to its
orbital decay. In order to verify this assumption we compute the
number of stars within a given distance from the centre of the system
and look for correlations between this quantity and $N_J$. In particular,
we define $N_r$ as the number of stars initially within a distance $r>r_h$
equal to the the hard binary separation which obtain $r<r_h$ at some
time during the simulation. For our fiducial choice
of mass ratios $M_{\rm bin} / M = 0.001, 0.005, 0.01$ this corresponds
to $r_h= 0.008, 0.014, 0.018$.

We find that $N_r$ correlates with $N_J$ in all models, both spherical
and non-spherical, as shown in figure \ref{fig:njnr}. We also find the
expected difference between spherical and triaxial models, with the
latter showing much larger values of $N_r$ than the corresponding
spherical models with the same $N$. The same holds for models T1 and
T2, where the more triaxial set systematically has larger values of
$N_r$. As usual, axisymmetric model lie in between the spherical and
the triaxial models.

\begin{figure}
  \begin{center}
    \includegraphics[height=8cm,angle=270]{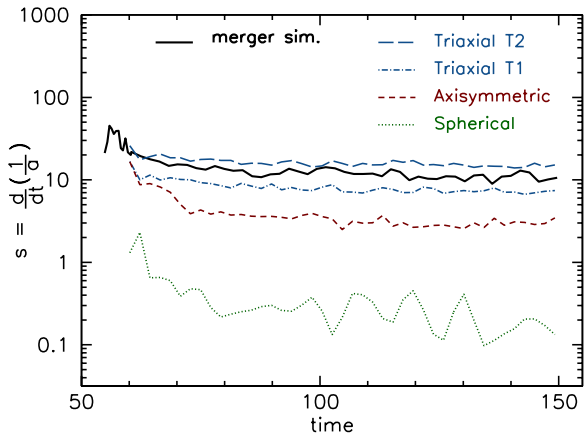}
  \end{center}
  \caption{Hardening rate in the spherical (dotted line), axisymmetric
    (short dashed line) and triaxial (dot-dashed and long dashed
    lines)  isolated models as a function of time. For a comparison,
    the hardening rate measured in the merger simulation with
    $\gamma=1$ and $N=10^6$ is also shown (thick solid line).}
  \label{fig:numhard}
\end{figure}

To determine the final fate of BHBs, we also computed the numerical
hardening rate $s_{NB}$ as given in equation~\ref{eq:hardnb},
assuming a starting value of the binary semi-major axis equal to the
one measured in the merger simulation with $\gamma=1$ and $N=10^6$.
In order to mimic the erosion of the stellar cusp produced by a BHB,
we assume that stars are lost after a time equal to their radial
orbital period and therefore do not contribute to the hardening rate
at later times.
The rates for spherical and non-spherical models with $N=10^6$ are
shown in Fig.~\ref{fig:numhard} as a function of time. 
We find that the triaxial models well reproduce the time evolution of
the hardening rates measured in the merger simulation, while the
spherical model shows an hardening rate that is more than one order of
magnitudes lower. The axisymmetric model is intermediate, with rates
significantly lower than the merger simulation.  The rates for the
merger simulation lie in between those of model T2, with axis ratios
$1:0.8:0.6$ and model T1, with axis ratios $1:0.9:0.8$. At the time of
binary formation, the merger remnant can be best fit by a triaxial
ellipsoid with axis ratios $1:0.8:0.6$ at a distance $D=0.5$ from the
binary's centre of mass, and by a triaxial ellipsoid with axis ratios
$1:0.9:0.7$ at a distance $D=1$. The shape of the remnant is therefore
consistent with an isolated model of triaxiality that is intermediate
between that of models T1 and T2. On the other hand, one might expect
the hardening of the merger remnant to be enhanced compared to that of
an isolated model due to rotation, which has been shown to lead to
higher hardening rates \citep{HBK2015}.  The small discrepancy might
be due to the fact that the simple semi-analytic approach used to
compute $s_{NB}$ assumes that any star with $\chi <1$ will interact
with the binary within an orbital period, and this might not always be
the case.

The hardening rate may also be artificially increased by Brownian
motion, a wandering of the binary which leads to an enlarged loss-cone
\citep{merritt2001, chat2003}.  Using dedicated $N$-body simulations,
however, \citet{bortolas2016} find that Brownian motion does not
affect the evolution of BHBs in simulations with $N$ in excess of one
million.

The timescale for coalescence from a separation $a$ due to emission of
gravitational waves is, for a circular binary,
\begin{eqnarray}
\label{eq:tgw}
T_{\rm GW} &=& \frac{5}{256} \frac{c^5 a^4}{G^3 M_1 M_2 \mbin} \nonumber\\
& \approx & 5.8\times 10^8 {\rm yr} \left(\frac{a}{10^{-3} \pc}\right)^4 \left(\frac{10^6 \msun}{\mbin}\right)^3\,.
\end{eqnarray}
Therefore, for a given binary mass the separation corresponding to a
time shorter than a Hubble time is
\begin{equation}
a_{\rm GW} \approx 2 \times 10^{-3} \pc \left(\frac{\mbin}{10^6 \msun}\right)^{3/4}.
\end{equation}
This corresponds to approximately $a_{\rm GW} \sim 0.01 a_h$ for all
considered models.  The total time to coalescence is given by the time
spent in hardening down to a separation of the order of $a_{\rm GW}$
plus the time $T_{\rm GW}$ for inspiral. For the models considered
above for the computation of $s_{\rm NB}$ with $\mbin = 10^6 \msun$
and extrapolating the evolution of $a(t)$ to later times, we find that both the
triaxial and axysimmetric models would be able to reach coalescence
within a Hubble time. However, for typical systems with $\mbin \sim
10^8 \msun$ only the triaxial models have hardening rates large enough
to ensure coalescence. The rates for the axisymmetric models are too
low to bridge the gap to the gravitational waves dominated
regime. These results are in good agreement with those of
\citet{VAM2015}.

\subsection{Effect of numerical softening}
\label{sec:softening}
The refilling parameter measured in the isolated models depends on the
numerical softening chosen in the simulations.  Softening reduces the
collisionality of a stellar system by suppressing the importance and
the effect of close encounters. Therefore it is natural to expect a
dependence of the refilling parameter on softening in spherical
systems, in which loss cone refilling is dominated by collisional
effects. In triaxial models, collisional and collisionless modes of
losscone repopulation coexist at the modest particle numbers typical
of direct summation simulations. If a small softening is used, as in
the simulations of \citet{VAM2014} including the BHB, very large
particle numbers are required for collisionless refilling to become
dominant. Hence their conclusion that much larger $N$ values than
affordable by direct summation are necessary to suppress relaxation
effects. On the other hand, adopting a larger value of softening, as
done in this work for models without MBHs, allows the effects of
global torques to become apparent at smaller particle numbers.

\begin{figure}
  \begin{center}
    \includegraphics[height=8cm, angle=270]{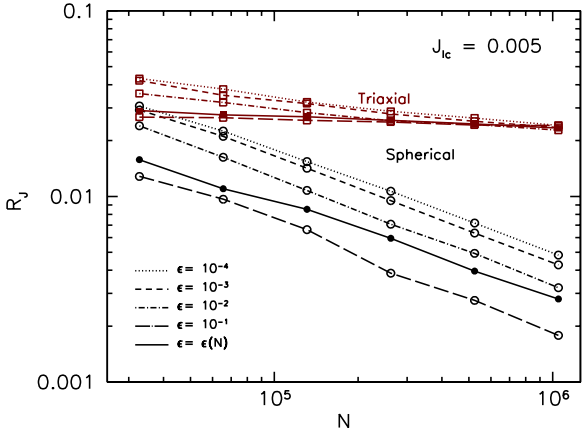}
  \end{center}
  \caption{Refilling parameter as a function of particle number in
    direct summation integrations of spherical (circles) and triaxial
    (squares) models with different values of softening. Empty symbols
    refer to fixed values of softening for all particle numbers, while
    filled symbols refer to the $N$-dependent prescription defined in
    equation \ref{eq:soft}.}
  \label{fig:soft}
\end{figure}

Figure~\ref{fig:soft} shows the dependence of $R_J$ on softening in
{\phiGRAPE} integrations of spherical and triaxial models. We
consider cases in which a fixed softening is adopted for models of
different $N$ (empty symbols) and the case of $N$-dependent softening
defined in equation \ref{eq:soft} (filled symbols).  In the spherical models $R_J$
decreases with increasing softening as expected, and the choice of the
variable softening works well at reducing collisional effects. In
triaxial models the effect of a larger softening is to reduce
scatterings into the loss cone and the flattening of $R_J$ appears at
smaller $N$. 

\section{Discussion and conclusions}
\label{sec:disc}
Black hole binary hardening in simulations of galaxy mergers is
believed to be sustained by a collisionless mode of loss cone
refilling which owes to global torques in non-spherical
potentials. This is necessary to ensure hardening down to separations
where emission of gravitational waves becomes dominant and leads the
black holes to coalescence. Merger remnants typically show a
significant degree of flattening and a modest departure from
axisymmetry \citep{preto2011,khan2011,GM2012}, which argues for a
significant population of stars on centrophilic orbits.  However, if
this is indeed the case then binary hardening should also be efficient
in isolated triaxial models. Simulations by \citet{VAM2014} show
that collisionless losscone refilling is masked by collisional
refilling due to stellar scatterings in simulations with modest
particle numbers and small softening. Here we take a different
approach to the problem and study losscone refilling in isolated
galaxy models. In order to reduce the effects of collisionality we
perform simulations with the fast multiple method code {\Griffin}
\citep{Dehnen2014}, which allows us to increase particle number to 64
million. 
We find a proxy for binary hardening, which is given by the refilling parameter, i.e. the
fraction of stars with angular momentum smaller than the angular
momentum of an hypothetical BHB of given mass, where each star is
counted only once per simulation.

Our key findings are:
\begin{itemize}
\item The refilling parameter, i.e. the fraction of stars which can be
  counted at least once to have angular momentum smaller than the loss
  cone angular momentum, is a good proxy for the hardening rate in
  merger simulations.
\item Loss cone refilling in spherical models depends critically on
  particle number, a clear signature that it is driven by two body
  scatterings. In real galaxies, this process becomes extremely
  inefficient and BHBs are not expected to reach the gravitational
  wave phase.
\item There is no $N$-dependence of the refilling parameter
  in triaxial models above $N \sim 10^7$. Refilling is more efficient in triaxial models
  than in spherical models, and a higher degree of triaxiality also
  leads to more efficient refilling.
\item Axisymmetric models have properties in between spherical and
  triaxial models. While refilling is consistently more efficient than
  in spherical cases, we observe a marked $N$-dependence with no
  obvious flattening even at the largest $N$ values.
\item Hardening rates computed directly from the $N$-body data in
 isolated triaxial models match those computed in merger simulations. On the
 other hand, spherical isolated models and axisymmetric models have
  significantly lower hardening rates. 
\item The hardening rates measured for the triaxial models are large
  enough to ensure coalescence of the binaries within a Hubble time
  for Milky Way type galaxies as well as more massive ones.  Hardening
  rates in axisymmetric models are only marginally sufficient to
  bridge the gap to the gravitational waves regime in the case of low
  mass binaries ($\mbin \lesssim 10^6 \msun$) but imply coalescence
  times that are longer than a Hubble time for typical binaries
  ($\mbin \sim 10^8 \msun$).  Spherical models are generally
  characterised by hardening rates too low to lead to coalescence.
\end{itemize}

\section*{Acknowledgments}
This work used the GPU cluster of the Astrophysics group, University
of Surrey, and the DiRAC Complexity system, operated by the University
of Leicester IT Services, which forms part of the STFC DiRAC HPC
Facility (www.dirac.ac.uk). This equipment is funded by BIS National
E-Infrastructure capital grant ST/K000373/1 and STFC DiRAC Operations
grant ST/K0003259/1. DiRAC is part of the National E-Infrastructure.
We thank Eugene Vasiliev for support with the SMILE software and
Andrew Pontzen for interesting discussions.

\bibliographystyle{mn2e}
\bibliography{biblio}

\appendix
\section{On the importance of force errors}

\begin{figure*}
  \begin{center}
    \includegraphics[height=8cm]{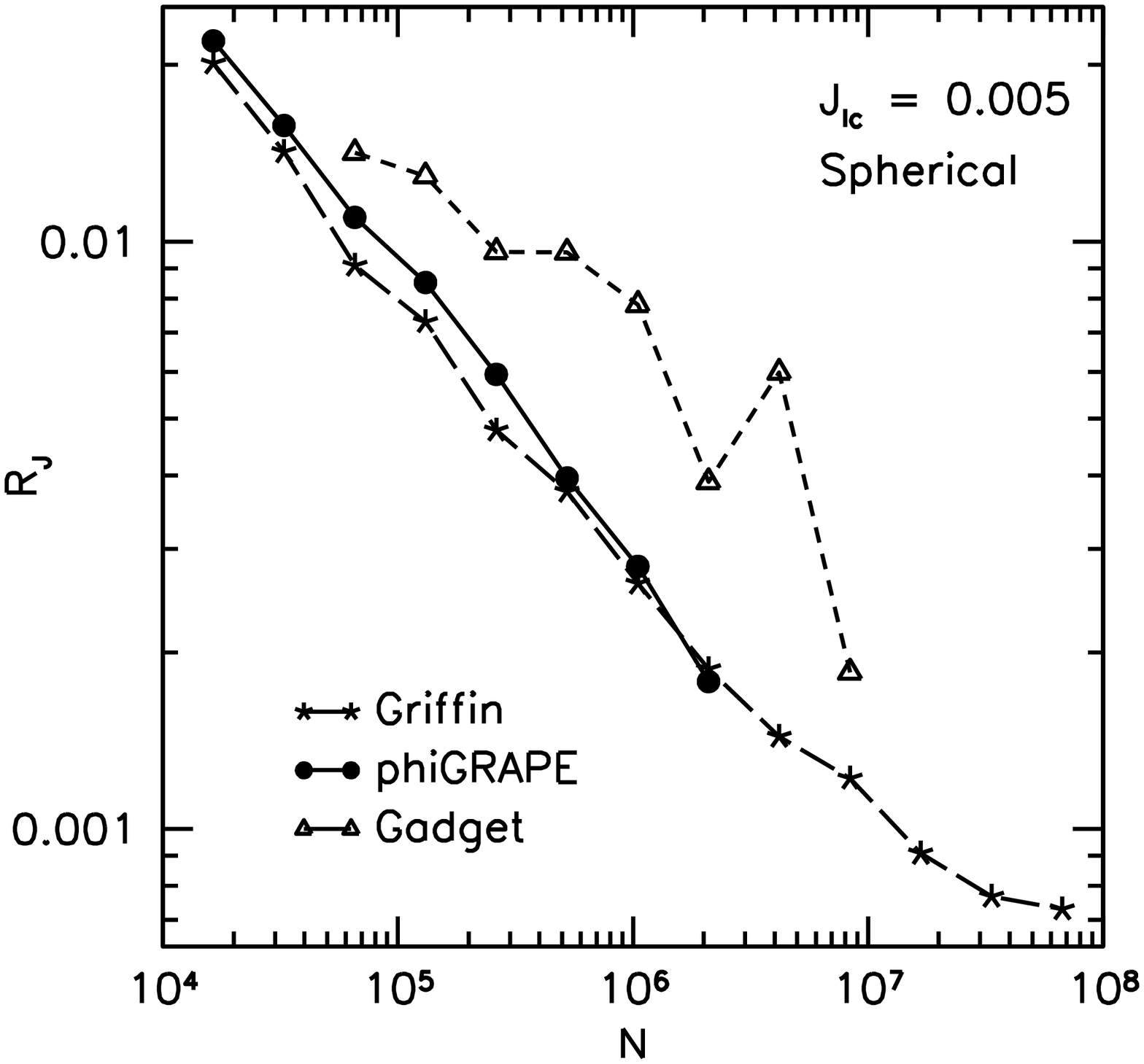}
    \includegraphics[height=8cm]{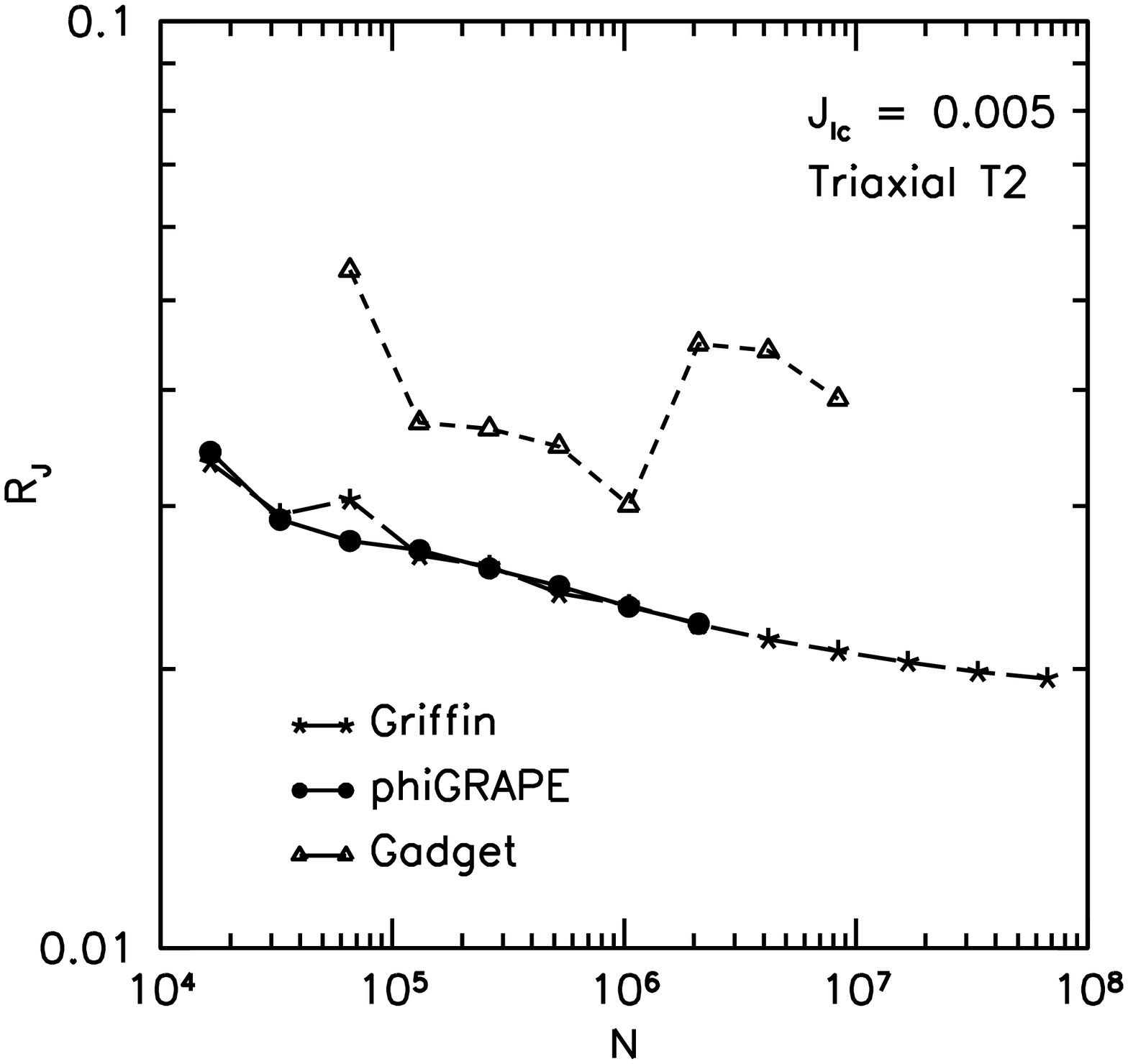}
  \end{center}
  \caption{Repopulation parameter versus particle number in spherical
    (left) and triaxial (right) models, evolved with three different
    codes: {\Gadget} (triangles), {\phiGRAPE} (circles) and {\Griffin} (stars).}
  \label{fig:gadgetsucks}
\end{figure*}
Moving from a direct summation force calculation to a faster
approximate method may seem a like natural step when trying to
increase particle number in order to reduce collisional
effects. However, we caution the reader that care needs to be taken
when choosing an approximate method for this type of problem. In order
to study diffusion in angular momentum driven by global torques, the
force errors need to be sufficiently small. \citet{Dehnen2014} presents an
analysis of the force errors for the fast multiple method (or tree
code) with a conventional geometric multipole-acceptance criterion
(based on an opening angle $\theta$). When using sufficiently large
expansion order $p$ and small $\theta$, the relative force errors of
such a method can be reduced to $10^{-7}$, the same level as
contemporary implementations of direct summation. However, the
approximated methods show extended tails towards large force errors, a
direct consequence of the simple geometric opening criterion. This
tail of a few stars with large force errors does not compromise global
energy conservation but may seriously affect the validity of
simulations, in particular if accurate representation of orbits is
required. Our method of choice, \Griffin, uses another type of opening
criterion, which is informed by an error estimate based on the
multipole moments themselves \citep[see][for details]{Dehnen2014}, and
produces a well-behaved distribution of force errors while minimising
the computational costs to obtaining a scaling better than $\mathcal{O}(N)$.

Fig.~\ref{fig:gadgetsucks} shows a comparison between \Gadget (a
conventional tree code with geometric opening criterion), \phiGRAPE
and \Griffin for the calculation of the refilling parameter in
spherical and triaxial models of varying $N$ (using the same
$N$-dependent softening for all codes). Values of $\rj$ agree
remarkably well between \phiGRAPE and \Griffin. For a
collision-dominated scenario (spherical), \Griffin errs slightly
towards too small $\rj$, which can be understood by a less accurate
time integration of two-body encounters, resulting in less
scattering. For the collisionless scenario (triaxial), \Griffin agrees
very well with \phiGRAPE, despite relative forces errors which are
$\sim 500$ times less accurate ($\times10^{-7}$ vs. $5\times10^{-5}$). The
  results from \Gadget are markedly different and do not reproduce the
  expected scaling with increasing $N$, but give too large $\rj$ with
  considerable scatter. This erroneous behaviour is exactly what one
  expects from a few large force errors, which act like a random
  relaxation process.

\end{document}